\documentclass{nature}
\usepackage{graphicx} 
\usepackage{cite}
\usepackage{epsfig} 
\usepackage{xcolor}
\usepackage{amsmath}
\usepackage{soul}
\usepackage{ulem}
\usepackage{physics}

\title{Polaritonic Chern insulators in monolayer semiconductors}

\author{Li He$^{1}$, Jingda Wu$^2$, Jicheng Jin$^1$, Eugene J. Mele$^1$, \& Bo Zhen$^{1,*}$}

\begin{document}
\maketitle

\begin{affiliations}
    \item Department of Physics and Astronomy, University of Pennsylvania, Philadelphia, PA 19104, USA \item Department of Physics and Astronomy, University of British Columbia, Vancouver, BC V6T 1Z1, Canada
\end{affiliations}

\begin{abstract}
Systems with strong light-matter interaction opens up new avenues for studying topological phases of matter. 
Examples include exciton-polaritons, mixed light-matter quasiparticles, where the topology of the polaritonic band structure arises from the collective coupling between matter wave and optical fields strongly confined in periodic dielectric structures. Distinct from light-matter interaction in a uniform  environment, the spatially varying nature of the optical fields leads to a fundamental modification of the well-known optical selection rules, which were derived under the plane wave approximation. 
Here we identify polaritonic Chern insulators by coupling valley excitons in transition metal dichalcogenides (TMDs) to photonic Bloch modes in a dielectric photonic crystal  (PhC) slab. 
We show that polaritonic Dirac points (DPs), which are markers for topological phase transition points, can be constructed from the collective coupling between valley excitons and photonic Dirac cones in the presence of both time-reversal and inversion symmetries.
Lifting exciton valley degeneracy by breaking time-reversal symmetry leads to gapped polaritonic bands with non-zero Chern numbers. 
Through numerical simulations, we predict polaritonic chiral edge states residing inside the topological gaps.
Our work paves the way to further explore polaritonic topological phases and their practical applications in polaritonic devices.\end{abstract}

\noindent\textbf{Introduction}

Strongly coupling photons to excitonic resonances in semiconductors gives rise to new elementary excitations, exciton-polaritons. 
These quasiparticles combine the properties of their parental photon and matter components and, thereby, provide an ideal platform to study quantum many-body physics, including Bose–Einstein condensates\cite{byrnes2014exciton}, superfluidity\cite{amo2009superfluidity}, and Berezinskii–Kosterlitz–Thouless transition\cite{kosterlitz1973ordering}. 
Motivated by the recent rapid progress in topological band theory\cite{hasan2010colloquium,lu2014topological,ozawa2019topological}, there has been of great interest in applying it to polaritonic systems and exploring novel topological phenomena intrinsic to exciton-polaritons\cite{karzig2015topological,bardyn2015topological,nalitov2015polariton,yi2016topological,klembt2018exciton,liu2020generation,li2021experimental}. 
For example, in contrast to photons that rarely interact with each other, exciton-polaritons exhibit strong nonlinearities through exciton-exciton interactions\cite{sun2017direct,togan2018enhanced,zhang2021van}, allowing for the study of nonlinear and interacting topological phenomena\cite{ozawa2019topological}. 
Another particularly promising perspective of exciton-polaritons is the possibility of breaking time-reversal symmetry ($T$) via the excitonic component, opening the door to the realization of a range of $T-$broken topological phases\cite{lu2014topological,he2020quadrupole} at optical frequencies, such as Chern insulators with protected unidirectional modes on the edges\cite{lu2014topological,ozawa2019topological}. 
By contrast, exploring such topological phenomena in purely photonic systems remains challenging due to the lack of efficient method to break $T$ for photons.

Recently, various topological exciton-polariton states have been theoretically proposed\cite{karzig2015topological,nalitov2015polariton,bardyn2015topological,yi2016topological} and experimentally studied in the solid-state setting by embedding semiconductor quantum wells into lattices of coupled micro-cavities\cite{klembt2018exciton}, or transferring two-dimensional TMDs onto dielectric photonic crystal slabs\cite{liu2020generation,li2021experimental}.
However, exciton-photon interaction in such periodic structures, which underlies the topology of the resulting polaritonic band structures, has neglected the periodic modulation of the refractive index and the spatially varying field strength and polarization\cite{lodahl2017chiral} --- a feature unique for strongly confined optical fields. The non-uniform distribution of polarization --- local transverse photon spin --- completely modifies the optical selection rules in light-matter interaction derived under the assumption of a plane wave excitation with a single polarization.  
As a result, a comprehensive understanding of the light-matter interaction in periodic structures is expected to change the topological properties of the exciton-polaritons in a fundamental way. 

Here, we present theoretical investigations of exciton-polaritons formed by strongly coupling monolayer TMDs to a photonic crystal slab, and show how polaritonic Chern insulators can be realized by breaking the time-reversal symmetry via the exciton components.
We start by deriving the effective Hamiltonian that captures the optical selection rules in periodic structures near high-symmetry momentum (\textbf{k}) points.
Based on that we show the polaritonic Dirac points can be constructed by coupling exciton resonances to photonic Dirac points. 
Similar to electronic and photonic systems, polaritonic Dirac points are also protected by the product of inversion ($P$) and time-reversal symmetry. As a result, breaking $T$ while preserving $P$ gaps out polariton Dirac points, leading to polaritonic bands with nontrivial Chern numbers. 
We validate the topological polaritonic band structures through finite-difference time-domain (FDTD) simulations.
All calculations are based on realistic material parameters\cite{li2014measurement,zhou2020controlling}, which makes our predictions immediately applicable for experimental realization.

\noindent\textbf{Results}

We start by introducing the concept of collective coupling between exciton resonances in a TMD monolayer and photonic Bloch modes in a dielectric photonic crystal slab.
One crucial feature of exciton-polariton systems is the significant size mismatch between the periodic potentials for photons and excitons (Fig.1a):
the photonic unit cell ($a \sim 300$ nm) is orders of magnitude larger than its excitonic counterpart ($a_X \sim 0.3$ nm). 
Accordingly, to study the exciton-polariton dispersions that are defined in the photonic Brillouin zone (BZ), one needs to fold the free exciton dispersion back into the photonic BZ, leading to highly degenerate excitonic bands with $N$-fold near degeneracy, where $N \gg 1$. Note that here due to the significant size mismatch, we neglect exciton band dispersion and assume it to be flat. 
The coupling between $N$ degenerate exciton modes and a single photon mode is described by the Tavis-Cummings Hamiltonian\cite{tavis1968exact}, a matrix of $N+1$ dimensions: 
\begin{equation}
    H_k = \omega_p(k) a_k^\dagger a_k + \sum_i^N \left( \omega_X b_{k,i}^\dagger b_{k,i} + g_i(k) a_k^\dagger b_{k,i} + g_i^*(k) b_{k,i}^\dagger a_k \right)  
\end{equation}
where $a_k^\dagger$ and $b_{k,i}^\dagger$ are creation operators for photon and $i$-th exciton mode with wavevector $k$ and eigen-energies $\omega_p$ and $\omega_X$, respectively. The coefficient $g_i$ denotes the coupling strength between the $i$-th exciton mode and the photon mode.
As shown in Fig.1b, in the strong coupling regime, exciton modes collectively couple to the photon mode, leading to two ``bright" polariton branches, namely the upper and lower polariton bands, with eigenvalues of $E_\pm = \frac{\omega_p + \omega_X}{2} \pm  \frac{1}{2} \sqrt{(\omega_p - \omega_X)^2 + 4 \sum_i^N |g_i|^2}$. 
In addition, there are $N-1$ ``dark" exciton modes at the original exciton energy $\omega_X$, residing inside the polariton gap, as they are completely decoupled from the photon mode. 
The fact that the photon mode only mixes with a linear superposition of exciton modes is a direct consequence of the degeneracy in exciton energies, similar to the super-radiance phenomenon in the context of quantum electrodynamics: an ensemble of non-interacting emitters couple to a common electromagnetic field through the formation of a collective mode\cite{gross1982superradiance}.
In general, when an ensemble of exciton states couple to $N_p$ photon modes, the number of ``bright" polariton bands is $N_{\text{polariton}} = N_p (N_X + 1)$, where $N_X$ is the number of non-degenerate exciton species. In the simplest example above, $N_X =1$, $N_p = 1$, and $N_{\text{polariton}} = 2$.

Next, we show how polaritonic Dirac points can be constructed by collectively coupling valley excitons in TMDs to photonic Dirac points. 
As a concrete example, we consider a Si$_3$N$_4$ PhC slab ($n=2.02$, thickness $230$ nm) with a honeycomb lattice of triangular holes (Fig.2a), which hosts photonic Dirac points at the $K/K'$ corners of the photonic BZ\cite{collins2016integrated}. 
The two degenerate photonic modes at the Dirac points can be characterized by their threefold rotation ($C_3$) eigenvalues $c_3^\pm = e^{\pm i 2\pi/3}$. 
For simplicity, we further assume that the energy of the TMD valley excitons, $\omega_{K/K'}$, is on-resonance with the photonic Dirac point $\omega_p$.
This is not an essential simplification since a small energy detuning will not qualitatively modify the polaritonic dispersion and the polariton Dirac points present below will remain (see Supplementary Note).
We can then represent the polariton wavefunctions using a six-component spinor $\Psi = \left( \ket{p^+}, \ket{K^+}, \ket{{K'}^+}, \ket{p^-}, \ket{K^-}, \ket{{K'}^-} \right) $ and write down the low-energy Hamiltonian near the $K$ point of the photonic BZ as follows: 

\begin{equation}
    H_\textbf{k} = 
    \begin{bmatrix}
    \omega_p &\alpha &\beta &v (k_x - i k_y) &0 &0 \\
    \alpha^* &\omega_K &0 &0 &0 &0 \\
    \beta^*  &0 &\omega_{K'} &0 &0 &0 \\
    v (k_x + i k_y) &0 &0 &\omega_p &\beta^* &\alpha^* \\
    0 &0 &0 &\beta &\omega_K &0 \\
    0 &0 &0 &\alpha &0 &\omega_{K'} 
    \end{bmatrix}
\end{equation}
Here $\textbf{k}$ is the wavevector measured relative to the photonic $K$ point. $\ket{p^\pm}$ are the two degenerate photon modes at the photonic Dirac point.
$\ket{K^{\pm}}$ and $\ket{K'^{\pm}}$ are the collective exciton modes formed by $K$ and $K'$ valley excitons with eigenvalues of $c_3^{\pm}$, respectively. 
Notably, the lack of continuous translation symmetry in PhCs gives rise to spatially varying field polarization in a photonic unit cell (Fig.2c), which in turn, leads to different coupling strengths ($|\alpha| \neq |\beta|$) between valley excitons and a fixed photon mode.
Moreover, neither of the valley excitons is completely decoupled from the photon mode as $|\alpha,\beta| \sim  \sqrt{\int_{\text{unit cell}} dx dy |E_{\pm}|^2} > 0$, where $E_{\pm} = (E_x \pm i E_y)/\sqrt{2}$ is the electric fields of $\ket{p^+}$ photon mode decomposed into the circular basis. 
This is to be distinguished from previous studies on semiconductor exciton-polaritons\cite{yi2016topological} and TMD-PhC systems\cite{li2021experimental,wang2020routing}, where neither of these important features are presented.

In the presence of time-reversal symmetry, 
$K$ and $K'$ valley excitons are degenerate in energy ($\omega_K = \omega_{K'}$ and $N_X$ = 1).
Consequently, the hybridization of excitons and photons generates $2 \times (1+1) = 4$ polariton bands that are doubly degenerate, forming upper and lower polaritonic Dirac cones at energies of $E_{\pm} = \omega_p \pm \sqrt{|\alpha|^2 + |\beta|^2}$. 
In addition to the four bright polariton bands, there are also two dark exciton bands with no photon components.
To confirm the existence of polaritonic DPs, we calculate polaritonic band structures using FDTD method, where the TMD monolayer is modelled as a dielectric material with Lorentz poles in its permittivity tensor $\bar{\bar{\varepsilon}}$ (see Supplementary Note for details).
The simulation result is shown in Fig.2d, which is in good agreement with the analytical result obtained from the effective Hamiltonian with the exciton-photon coupling strength ($|\alpha|$ and $|\beta|$) being fitting parameters.  
The observed Rabi splitting between the two DPs is $ 2 \sqrt{|\alpha|^2 + |\beta|^2} = 28$ meV, which is comparable to experimental results obtained in similar settings of TMD-PhC polaritons\cite{zhang2018photonic,liu2020generation,li2021experimental}. 

Similar to DPs in two-dimensional electronic and photonic systems, the presence of polaritonic DPs are protected by the product of inversion and time-reversal symmetries.
Therefore, breaking $P$ or $T$ lifts the degeneracies, leading to gapped phases with distinct bulk topologies.
To break $P$, one can allow the two holes in a photonic unit cell to have different radii. 
This is equivalent to adding a mass term $m\sigma_z$ to the bare photonic Dirac cone, and, consequently, gapping out both polaritonic Dirac points (Fig.3a).
However, due to the presence of time-reversal symmetry, the resulting energy gaps are topologically trivial with zero Chern numbers ($C$), as can be confirmed by using the $C_3$ eigenvalues at high-symmetry $\mathbf{k}$ points of all bands below the energy gaps\cite{fang2012bulk}: $e^{i2\pi C/3} = \prod_i c_{3,i} (\Gamma) c_{3,i} (K) c_{3,i} (K')$, where $c_{3,i}(K) = c_{3,i}^*(K')$ as required by $T$.

In contrast, the introduction of $T$-breaking perturbations by lifting the exciton valley degeneracy ($\omega_K \neq \omega_{K'}$ and $N_X = 2$) leads to distinct gapped energy spectrum with nontrivial bulk topology. 
One method to break $T$ is to apply a perpendicular magnetic field that induces a valley Zeeman splitting\cite{srivastava2015valley,aivazian2015magnetic,cunningham2019resonant}.
Another effective approach is through valley-selective optical stark effect, where large valley exciton energy splitting up to tens of meV has been observed experimentally even at room temperature\cite{sie2015valley,kim2014ultrafast,yong2019valley}.
When $T$ is broken, the block degeneracy in the effective Hamiltonian is lifted, and all bands hybridize together to yield $2\times(2+1)=6$ polaritonic bands (Fig.3b). 
This is in contrast to the $P-$breaking case where only four polariton bands are observed. 
Moreover, the presence of inversion symmetry relates the $C_3$ eigenvalues at $K$ and $K'$: $c_{3,i}(K) = c_{3,i}(K')$, 
This indicates that all the polariton bands have non-zero Chern numbers, which can be inferred directly from the phase winding in the $H_z$ mode profiles at $K$ point (Fig.3c). 
Remarkably, the observed topological gap size of $4.1$ meV induced by a TMD monolayer is at least an order of magnitude larger than that of semiconductor quantum well exciton polaritons\cite{klembt2018exciton}, owing to the strong light-matter interactions in 2D TMDs. 
The topological phase diagrams for different energy gaps (shaded in gray) as a function of $P-$ and $T-$breaking strengths are shown in Fig.3d and e. 
The distinct phase transition boundaries of the middle gap (Fig.3d) as compared to the that of the bottom gap (Fig.3e) is due to the fact that the polariton bands above/below the middle gap are dominated by exciton component, which are not affected by $P-$breaking perturbations. 

Finally, we show the existence of polaritonic chiral edge states at the interfaces between a polaritonic Chern insulator and normal insulators. 
To this end, we consider a supercell geometry where a polaritonic Chern insulator is sandwiched between normal insulators that support a wide band gap for the energy range of interest.
To further enhance the topological gap size and hence the visibility of the chiral edge states in our simulation, we increase the exciton density by considering multiple TMD monolayers couple to the PhC slab\cite{dufferwiel2015exciton,gu2019room} (the exciton-photon coupling strength and the resulting topological gap size is proportional to $\sqrt{N_\text{layer}}$, see Supplementary Note for details). 
Fig.4b shows the projected dispersion of the semi-infinite strip. 
Edge states emerge at both top and bottom interfaces with their energies traversing across all the three bulk topological gaps, which indicates the unidirectional nature of the edge modes.

\noindent\textbf{Discussion}

In summary, we present a practical design of exciton-polariton Chern insulators by integrating TMD monolayers onto a dielectric photonic crystal slab. 
Using numerical calculations with realistic material parameters, we show large polaritonic topological gaps can be opened as a result of the strong light-matter interactions in 2D TMDs, featuring polaritonic chiral edge states inside the topological gaps. 
The predicted polaritonic chiral edge states can be readily observed by combining high quality TMD monolayers\cite{zhou2020controlling} with low-loss dielectric PhC slabs, which may enable novel polaritonic devices with topological protection, such as polaritonic topological lasers, isolators, circulators, and unidirectional amplifiers. 
The presented theoretical analysis of collective coupling between excitons and photonic Bloch modes is applicable to a wide range of material platforms such as quantum well microcavity exciton polaritons, and provides a comprehensive understanding of strongly coupled light-matter systems.
Finally, combining with the strong nonlinearities of excitons in TMDs\cite{zhang2021van}, TMD-PhC exciton-polaritons may provide a fertile experimental test bed to explore nonlinear and interacting topological phenomena beyond single particle level. 









\noindent\textbf{References}
\bibliography{reference}{}
\bibliographystyle{naturemag}

\begin{addendum}
\item[Methods]
\textbf{Numerical calculation of polaritonic band dispersions using finite-difference time-domain simulations.} The band structures and mode profiles are calculated using finite-different time-domain simulations in Lumerical 2021. Detailed simulation parameters are presented in Supplementary Note.

\item[\textcolor{black}{Data availability}]
\textcolor{black}{The data that support the findings of this study are available from the corresponding author upon reasonable request.}

\item[Acknowledgements]
We thank Jeff Young for support in FDTD simulations. This research was enabled in part by support provided by Westgrid and Compute Canada.
This work was partly supported by the National Science Foundation through the University of Pennsylvania Materials Research Science and Engineering Center DMR-1720530, the US Office of Naval Research (ONR) Multidisciplinary University Research Initiative (MURI) grant N00014-20-1-2325 on Robust Photonic Materials with High-Order Topological Protection and grant N00014-21-1-2703. EJM's work was supported by Department of Energy under grant DE-FG02-84ER45118.
 
\item[Author contributions] 
L.H. and B.Z. conceived the idea; L.H. developed the theory and carried out numerical simulations with the help of J.W.; All authors discussed the results; L.H and B.Z. wrote the paper with contributions from all authors; B.Z. supervised the project.

\item[Competing Interests] 
The authors declare no competing interests.
\item[Correspondence] Correspondence and requests for materials should be addressed to Bo Zhen.~(email: bozhen@sas.upenn.edu).
\end{addendum}

\clearpage
\begin{figure}
\centering
\includegraphics[width=8.8cm]{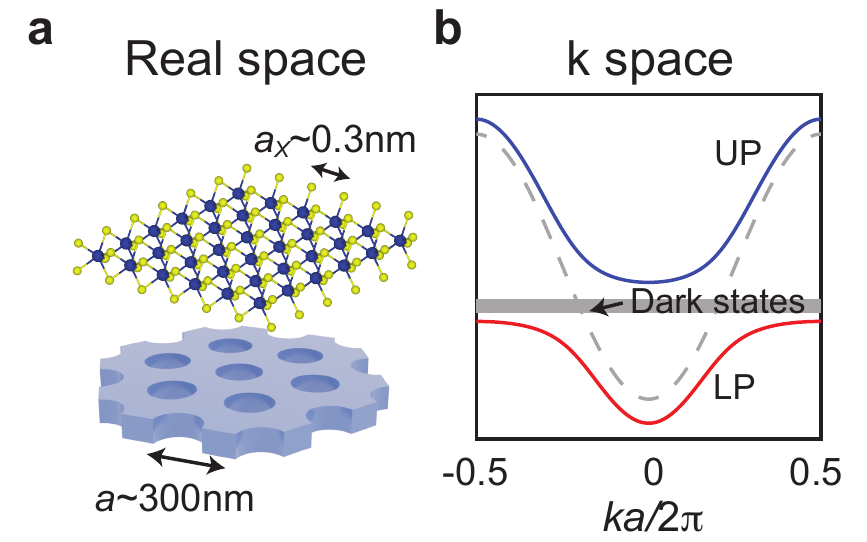}
\caption{{\bf Collective exciton-photon coupling in the TMD-PhC system $\mid$ }
{\bf a.} Schematics of a TMD monolayer placed on top of a dielectric PhC slab.
{\bf b.} Upper (blue) and lower (red) polariton branches arise from the collective coupling between a photon band (dashed parabola) and $N$-fold degenerate exciton bands. There are also $N-1$ exciton bands (shaded in gray) in the polariton gap that are decoupled from the photon band. 
}
\end{figure} 

\clearpage
\begin{figure}
\centering
\includegraphics[width=16cm]{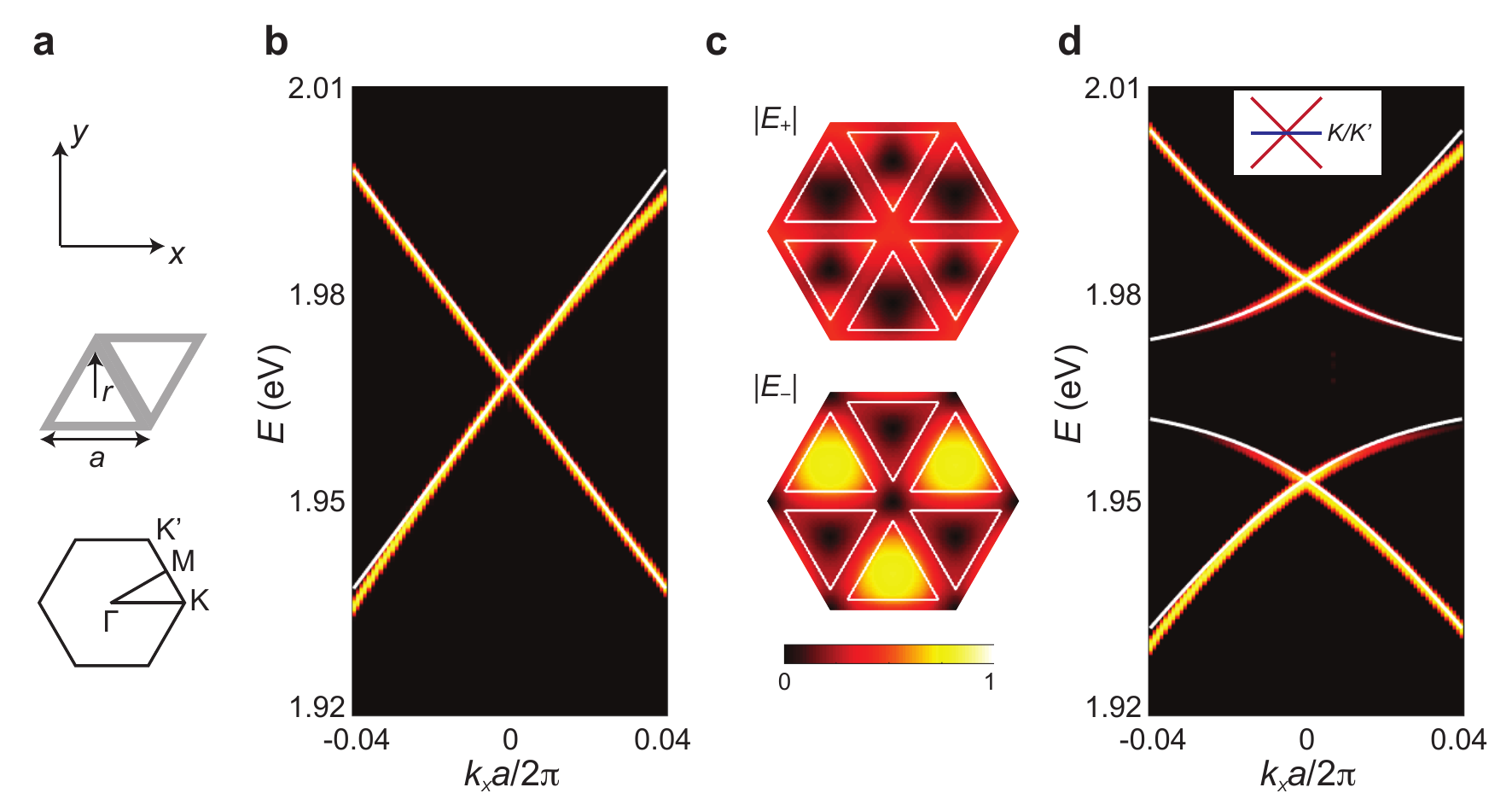}
\caption{
{\bf  Polaritonic Dirac points resulting from coupling valley excitons to a photonic Dirac point $\mid$}
{\bf a.} Si$_3$N$_4$ PhC slab with a honeycomb lattice of triangular holes (lattice constant $a$ = 310 nm, hole radius $r$ = 130 nm).
{\bf b.} Calculated photonic band structure showing a Dirac cone near the $K$ point.
{\bf c.} 
Decomposition of the $\ket{p^+}$ photon mode into two circular polarizations of $E_{\pm} = (E_x \pm i E_y)/\sqrt{2}$. 
The in-equivalent distributions of $|E_{\pm}|$ implies distinct coupling strengths between $\ket{p^+}$ and excitons in $K/K'$ valleys.
{\bf d.} Exciton-photon coupling generates a pair of polaritonic Dirac cones. White lines are the analytical results obtained from the effective Hamiltonian.
}
\end{figure} 

\clearpage
\begin{figure}
\centering
\includegraphics[width=16cm]{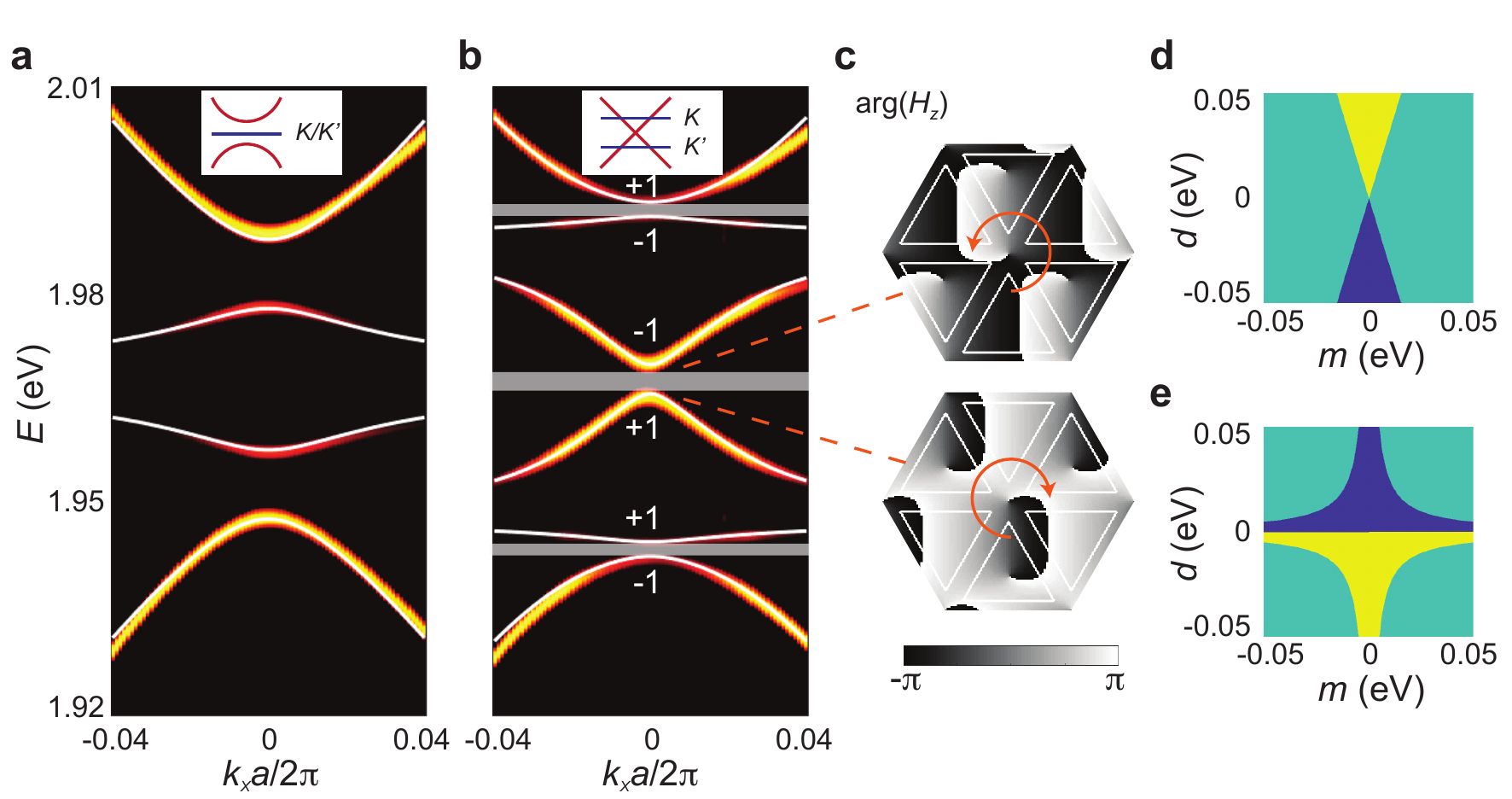}
\caption{{\bf Polaritonic topological phase transitions by lifting exciton valley degeneracy $\mid$ }
{\bf a,b.} Breaking either $P$ or $T$ removes the degeneracies at $K/K'$, leading to gapped polaritonic band structures with distinct bulk topology. In the $P-$breaking case ({\bf a}), the two holes in a unit cell have different radii ($r_1$ = 125 nm and $r_2$ = 135 nm). 
In the $T-$breaking case ({\bf b}), the exciton energies in $K/K'$ valleys are shifted by $d=\pm$19.7 meV with respect to $\omega_p$.
{\bf c.} 
$H_z$ mode profiles for the third and fourth bands at $K$ feature opposite phase winding of $\pm 2\pi$. 
{\bf d,e.} Topological phase diagrams for the first and third bulk energy gaps (shaded in gray) exhibit different phase transition boundaries. Yellow: $C=1$; blue: $C=-1$; green: $C=0$. 
}
\end{figure} 

\clearpage
\begin{figure}
\centering
\includegraphics[width=16cm]{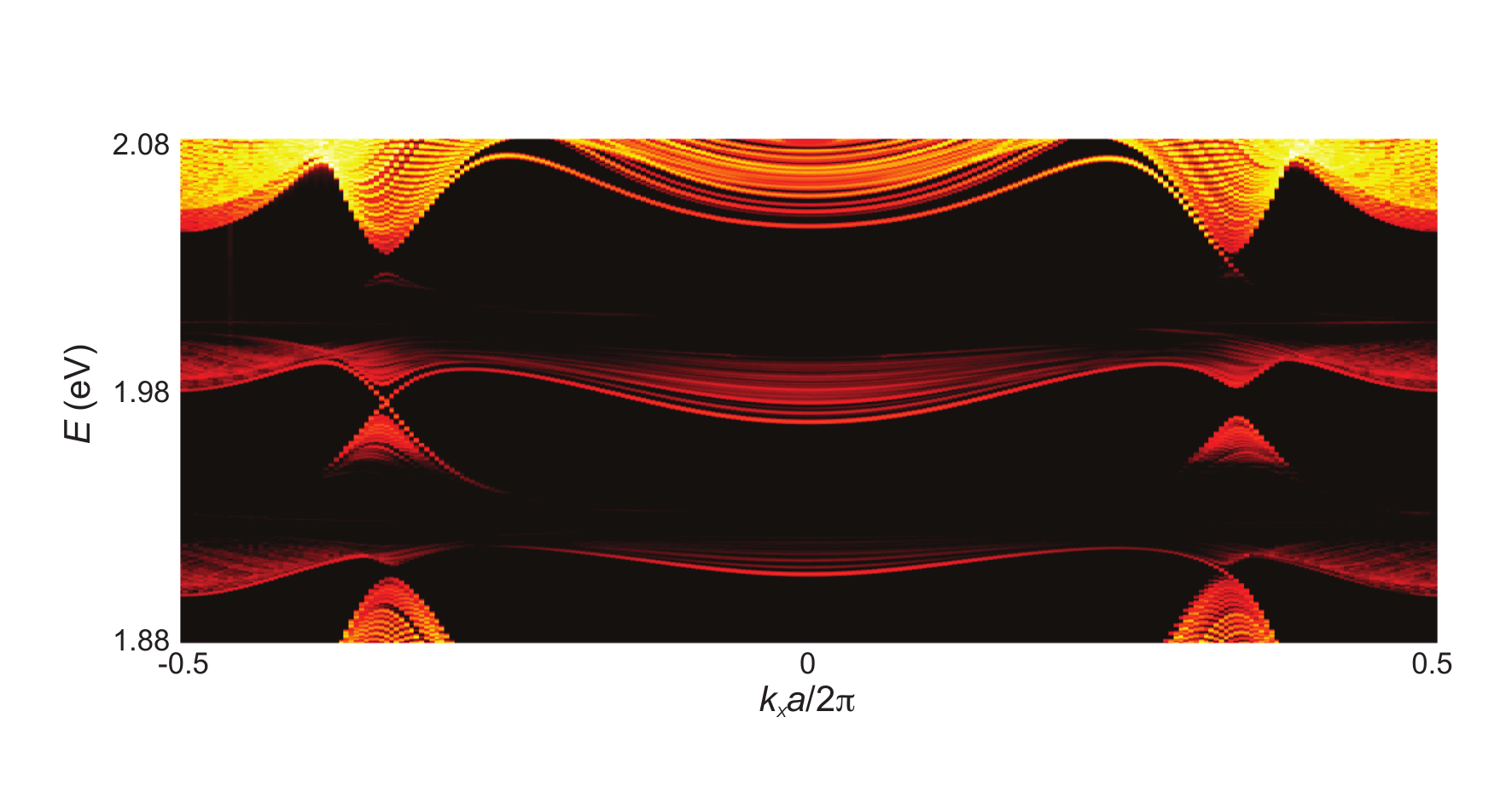}
\caption{{\bf $\mid$ Chiral edge states at the interfaces between a polaritonic Chern insulator and normal insulators.}
The normal insulators have similar design of the $P-$breaking case in Fig.3a, with $r_1 = 75$ nm and $r_2 = 165$ nm. 
Further details of the super-cell setup can be found in Supplementary Note. }
\end{figure}

\end{document}